\def\nk{n_{\rm b}}

\def\rfr#1{Equation\,(\ref{#1})}
\def\rfrs#1#2{Equations\,(\ref{#1})--(\ref{#2})}

\def\derp#1#2{\rp{\partial{#1}}{\partial{#2}}}

\def\virg#1{``#1"}

\def\eqi{\begin{equation}}
\def\eqf{\end{equation}}
\def\eqia{\begin{eqnarray}}
\def\eqfa{\end{eqnarray}}

\def\rp#1#2{\frac{#1}{#2}}
\def\lb#1{\label{#1}}

\def\bds#1{\boldsymbol{#1}}

\def\ton#1{\left(#1\right)}
\def\qua#1{\left[#1\right]}

\documentclass[onecolumn]{aastex}

\usepackage{morefloats}
\usepackage[title]{appendix}
\usepackage{textcomp}
\usepackage{booktabs}
\usepackage[table,xcdraw]{xcolor}
\usepackage{multirow}
\usepackage{rotating,tabularx}
\usepackage{float}
\usepackage{enumerate}
\usepackage{rotating}
\usepackage[polutonikogreek,english]{babel}
\usepackage{amsmath,starfont,textgreek,w-greek,wasysym}
\usepackage[flushleft]{threeparttable}
\usepackage{amsthm}
\usepackage{amscd,lineno}
\usepackage{amssymb,dsfont}
\usepackage{graphicx,epsfig}
\usepackage{txfonts}
\usepackage{xr-hyper}
\usepackage{hyperref}

\RequirePackage{color}

\newcommand{\emaila}{lorenzo.iorio@libero.it}

\allowdisplaybreaks[1]

\begin{document}

\title{Limitations in testing the Lense-Thirring effect with LAGEOS and \textcolor{black}{the} newly launched geodetic satellite \textcolor{black}{LARES 2}}

\shortauthors{L. Iorio}

\author{Lorenzo Iorio\altaffilmark{1} }
\affil{Ministero dell' Istruzione e del Merito
\\ Viale Unit\`{a} di Italia 68, I-70125, Bari (BA),
Italy}

\email{\emaila}

\begin{abstract}
\textcolor{black}{The} new Earth's geodetic satellite \textcolor{black}{LARES 2}, cousin of LAGEOS and sharing with it almost the same orbital parameters apart from the inclination, displaced by 180 deg, was launched last year. Its proponents suggest to use the sum of the nodes  of LAGEOS and of \textcolor{black}{LARES 2} to measure the sum of the Lense-Thirring node precessions independently of the systematic bias due to the even zonal harmonics of the geopotential, claiming a final $\simeq 0.2$ per cent total accuracy. In fact, the actual orbital configurations of the two satellites do not allow \textcolor{black}{one} to attain the sought for mutual \textcolor{black}{cancellation} of their classical node precessions due to the Earth's quadrupole mass moment, \textcolor{black}{as their sum is still} $\simeq 5\,000$ times larger than the added general relativistic rates. This has important consequences. One is that the current uncertainties in the eccentricities and the inclinations of both satellites do not presently allow the stated accuracy goal to be met, needing improvements by $3-4$ orders of magnitude. Furthermore, the imperfect knowledge of the Earth's angular momentum $S$ impacts the uncancelled sum of the node precessions,  from 150 to $4\,900$ per cent of the relativistic signal depending on the uncertainty assumed in $S$. It is finally remarked that the real breakthrough in reliably testing the gravitomagnetic field of the Earth would consist in modeling it and simultaneously estimating one or more dedicated parameter(s)  along with other ones characterizing the geopotential, as is customarily performed  for any other dynamical feature.
\end{abstract}



\section{Introduction}
To the first post-Newtonian (1pN) level, the gravitational field of an isolated, slowly rotating body of mass $M$ has a non-central, magnetic-like component, dubbed as \virg{gravitomagnetic} and sourced by its angular momentum\footnote{In the following, its orientation in space will be assumed to be known with sufficient accuracy, as in the case of the Earth, so that a coordinate system  with, say, the reference $z$ axis aligned with it will be adopted.} $\bds S$.

Actually, such a denomination has nothing to do with electric charges and currents; it is only due to the formal resemblance of the linearized equations of the General Theory of Relativity (GTR), in its weak-field and slow-motion approximation, with the Maxwell equations of electromagnetism. \textcolor{black}{In this general relativistic framework, the paradigm of \virg{gravitoelectromagnetism} arose \citep{1958NCim...10..318C,Thorne86,1986hmac.book..103T,1988nznf.conf..573T,1991AmJPh..59..421H,
1992AnPhy.215....1J,2001rfg..conf..121M,2001rsgc.book.....R,Mash07,2008PhRvD..78b4021C,
2014GReGr..46.1792C,2021Univ....7..388C,2021Univ....7..451R}.}
\textcolor{black}{It encompasses a series of entirely gravitational phenomena affecting orbiting test particles, precessing gyroscopes, moving clocks and atoms, and propagating electromagnetic waves \citep{1977PhRvD..15.2047B,1986SvPhU..29..215D,2002EL.....60..167T,2002NCimB.117..743R,2004GReGr..36.2223S,2009SSRv..148...37S}.}

\textcolor{black}{
General relativistic gravitomagnetism  should play a major role in several complex  processes which take place near spinning black holes and involve accretion disks and relativistic jets \citep{1975ApJ...195L..65B,1978Natur.275..516R,1982MNRAS.198..345M,1984ARA&A..22..471R,1988nznf.conf..573T,1999ApJ...525..909A,
2009MNRAS.397L.101I,2009SSRv..148..105S,2013ApJ...778..165V,2016MNRAS.455.1946F}. Also various hypothesized effects like the Penrose Process \citep{2002GReGr..34.1141P,1971NPhS..229..177P,2021Univ....7..416S}, the Blandford-Znajek effect \citep{1977MNRAS.179..433B} and superradiance \citep{1971JETPL..14..180Z} are attributable to the gravitomagnetic field of a rotating black hole; see \citet{2015CQGra..32l4006T} and references therein. Thus, it is important to experimentally check such a consequence of GTR  in as many different scenarios as possible in a reliable way in order to trustworthily extrapolating its validity also to other realms in which it is much more difficult and uncertain testing it.
}

\textcolor{black}{In particular}, the gravitomagnetic field of \textcolor{black}{a localized} spinning mass causes the orbital plane $\Pi$ of a test particle circling it to secularly precess independently of the inclination $I$ of $\Pi$ to the body's equator according to
\eqi
\dot\Omega_\mathrm{LT} = \rp{2\,G\,S}{c^2\,a^3\,\ton{1-e^2}^{3/2}};\lb{OLT}
\eqf
this is known as the Lense-Thirring (LT) effect \citep{1918PhyZ...19..156L,1984GReGr..16..711M}, \textcolor{black}{despite recent studies \citep{2007GReGr..39.1735P,2008mgm..conf.2456P,Pfister2014} showing that it should be more appropriately renamed as Einstein-Thirring-Lense effect}. In \rfr{OLT}, $\Omega$ is the satellite's longitude of the ascending node which determines where $\Pi$ intersects the body's equatorial plane,  $c$ is the speed of light in vacuum, $G$ is the Newtonian constant of gravitation,  $a,\,e$ are the satellite's semimajor axis and eccentricity, respectively.
In fact, the argument of pericentre $\omega$ of the test particle also undergoes a secular LT precession \citep{1918PhyZ...19..156L,1984GReGr..16..711M}
\eqi
\textcolor{black}{\dot\omega_\mathrm{LT} = -\rp{6\,G\,S\,\cos I}{c^2\,a^3\,\ton{1 - e^2}^{3/2}};}\lb{periLT}
\eqf
 it is not treated here since such an orbital element turns out to be heavily perturbed by several non-gravitational perturbations \citep{Nobilibook87}\textcolor{black}{, especially for geodetic, passive spacecraft whose orbit is almost circular \citep{2001P&SS...49..447L,2002P&SS...50.1067L}. Instead, the node is rather insensitive to the non-conservative accelerations \citep{1981CeMec..25..169S,2001P&SS...49..447L,2017AcAau.140..469P}.  It is worth mentioning that, for an arbitrary orientation of the spin axis $\bds{\hat{k}}$ of the central body in space, the inclination  experiences an  LT precession $\dot I_\mathrm{LT}$ \citep{1988NCimB.101..127D}; it vanishes in a coordinate system an axis of which is aligned with the spin of the rotating mass.}

A major competing effect of classical origin is caused by the even zonal harmonic coefficients $J_\ell,\,\ell=2,\,4,\,\ldots$ of the multipolar expansion of the Newtonian gravitational potential of the source body accounting for its departures from spherical symmetry \citep{Capde05}
\eqi
\textcolor{black}{U = -\rp{G\,M}{r}\,\qua{1 - \sum_{\ell = 2}^\infty\,J_\ell\,\ton{\rp{R}{r}}^\ell\,\mathcal{P}_\ell\ton{\bds{\hat{k}}\bds\cdot\bds{\hat{r}}}}.}\lb{Upot}
\eqf

\textcolor{black}{In \rfr{Upot},  $R$ is the body's equatorial radius, while $\mathcal{P}_\ell\ton{\cdots}$ is the Legendre polynomial of degree $\ell$ whose argument is the cosine of the angle between  $\bds{\hat{k}}$ and the position vector $\bds r$ of the test particle. }
 Indeed, the even zonals induce secular node precessions $\dot\Omega_{J_\ell}$ whose nominal values are usually several orders of magnitude larger than the LT ones\textcolor{black}{; for a calculation of $\dot\Omega_{J_\ell}$ up to the degree $\ell = 20$, see  \citet{2003CeMDA..86..277I}.} The largest one is  due to the first even zonal harmonic $J_2$; this is \citep{Capde05}
\eqi
\dot\Omega_{J_2} \lb{OJ2} = -\rp{3}{2}\,\nk\,J_2\,\ton{\rp{R}{a}}^2\,\rp{\cos I}{\ton{1-e^2}^2},
\eqf
where $\nk\doteq\sqrt{GM/a^3}$ is the satellite's mean motion.

From \rfr{OJ2}, it can be noted that $\dot\Omega_{J_2}$ depends on $I$ in such a way that for two satellites A and B ideally with the same orbital elements, apart from their orbital planes being exactly inclined $180\,\mathrm{deg}$ one from each other, the sum of their classical node precessions $\dot\Omega_{J_2}^\mathrm{A} + \dot\Omega_{J_2}^\mathrm{B}$ cancels out, while the LT ones add up. \textcolor{black}{The same holds also for the higher degrees since it turns out that $\dot\Omega_{J_\ell},\,\ell = 4,\,6,\,8,\ldots$ are proportional just to $\dot\Omega_{J_2}$ through linear functions of even powers of $\sin I$ \citep{2003CeMDA..86..277I}; as an example, for $\ell = 4$ one has \citep{2003CeMDA..86..277I}}
\eqi
\textcolor{black}{\dot\Omega_{J_4} = \dot\Omega_{J_2}\,\rp{J_4}{J_2}\,\qua{\rp{5}{8}\,\ton{\rp{R}{a}}^2\,\rp{1 + \rp{3}{2}\,e^2}{\ton{1 - e^2}^2}\,\ton{7\,\sin^2 I - 4}}.}
\eqf

\textcolor{black}{Such a distinctive property of the classical and LT node precessions was remarked on, for the first time, by \citet{1976CeMec..13..429V,1976PhRvL..36..629V} in the case of two counter-orbiting spacecraft with the same orbital parameters; it is essentially the same as for the previously mentioned \virg{butterfly} orbital configuration, or critical supplementary orbit configuration, in which the inclinations of the two satellites are displaced by $180\,\mathrm{deg}$, all the other orbital elements being the same. \citet{1976CeMec..13..429V,1976PhRvL..36..629V} considered the sum of the nodes of both their counter-revolving satellites to be endowed with drag-free apparatus, counterbalancing the disturbing non-gravitational accelerations.}
\citet{2023EPJC...83...87C}\textcolor{black}{, referring to an earlier proposal \citep{1986PhRvL..56..278C} equivalent to the strategy proposed by \citet{1976CeMec..13..429V,1976PhRvL..36..629V},} intend to exploit \textcolor{black}{the previously mentioned feature of the Newtonian and relativistic node rates} to accurately measure the sum of the LT node precessions of the  Earth's geodetic satellites LAGEOS (L) and of \textcolor{black}{LARES 2} (\textcolor{black}{LR 2}), both tracked with the Satellite Laser Ranging (SLR) technique \citep{SLR11}. \textcolor{black}{It is worth noticing that the idea of using artificial satellites to measure the LT effect in the field of the Earth dates back to the pioneering works by \citet{Ginz57a,Bogo59,Ginz59}, while using L and other SLR satellites existing at the time  was proposed for the first time by \citet{1978A&A....69..321C}.}  \textcolor{black}{\citet{2023EPJC...83...87C} state} to be able to finally reach a total accuracy as good as $\simeq 0.2$ per cent.
Ciufolini and coworkers have started to use L, put into orbit in 1976, and other laser geodetic satellites of the same kind \citep{2019JGeod..93.2181P} launched over the following years (LAGEOS 2, LARES) to measure the LT effect since 1996 \citep{1996NCimA.109..575C}, always \textcolor{black}{claiming} accuracies which were often questioned by subsequent studies by other authors \citep{2011Ap&SS.331..351I,2012CaJPh..90..883R,2013AcAau..91..141I,2013CEJPh..11..531R,2013NewA...23...63R,2014NewA...29...25R}; such a controversy is still lingering; see, e.g., \citet{2013CEJPh..11..531R} and references therein. \textcolor{black}{It may be interesting to note that, at least in principle, the critical supplementary orbit configuration would work well by looking at the perigee as well. Indeed, in this case, the classical apsidal shifts are identical \citep{2003CeMDA..86..277I}, while the LT ones are equal and opposite in sign, as per \rfr{periLT}. Thus, the difference of the perigee precessions would allow to ideally cancel out the bias due to the geopotential and, at the same time, to add up the LT rates \citep{2003GReGr..35.1583I}. However, such a proposal may not be effective for geodetic, passive spacecraft because of the impact of the non-gravitational perturbations sensibly affecting the perigee, contrary to the node.}

A so far unquestioned measurement of another gravitomagnetic effect in the field of the Earth with a quite expensive, dedicated spaceborne experiment was performed by the Gravity Probe B (GP-B) mission \citep{Varenna74} whose timeframe, ranging from its early conception to the release of its final results, lasted for about 40 years \textcolor{black}{at a cost of about \$ 750 million \citep{2011PhyOJ...4...43W}} . It measured the Pugh-Schiff precessions \citep{Pugh59,Schiff60} of the axes of four gyroscopes carried onboard to an accuracy of the order of 19 per cent  \citep{2011PhRvL.106v1101E}, despite its originally expected level \textcolor{black}{of} about $1$ per cent \citep{2001LNP...562...52E}. For an overview of other proposed LT tests with natural or artificial bodies in the Solar System, see \citet{2011Ap&SS.331..351I} and references therein.
\textcolor{black}{
A successful detection of the gravitomagnetic orbital precession of the inclination of the binary system PSR J1141-6545 \citep{2011MNRAS.412..580A}, made of a white dwarf and a pulsar,  was recently claimed \citep{2020Sci...367..577V}; later analysis raised concerns about such a test \citep{2020MNRAS.495.2777I}. Attempts to measure  the gravitomagnetic periastron precession of the double pulsar PSR J0737-3039A/B \citep{2003Natur.426..531B,2004Sci...303.1153L} in the next future are underway  \citep{Kehletal017,2020MNRAS.497.3118H}.
Evidence for manifestations of the gravitomagnetic field in the strong-field regime was often claimed over the years. Gravitomagnetism could be responsible of the quasi-periodic oscillations in X-ray binaries \citep{1998ApJ...492L..53C,1998ApJ...507..316M,1998ApJ...492L..59S}.
The Lense-Thirring precession induced by a slowly rotating compact object could be compatible with the daily variations of the ejecta angle observed in the microquasar $\mathrm{LS}\,\mathrm{I}+61^\circ 303$ \citep{2010A&A...515A..82M}. It was recently reported that the observed quasi-periodic modulation of the iron line centroid energy in the microquasar H1743-322 \citep{1977IAUC.3099....3K} may be produced by the Lense–Thirring effect \citep{2016MNRAS.461.1967I}. In general, the reliability and accuracy of such tests is difficult to be reliably assessed because of the relatively poor knowledge of the astrophysical environments in which the phenomena of interest take place.
}

In this paper, it is shown that the claims by \citet{2023EPJC...83...87C} \textcolor{black}{of} $\simeq 0.2$ per cent accuracy allegedly obtainable with L and \textcolor{black}{LR 2} are definitely unrealistic. In Section\,\ref{sec2} it is demonstrated that the sum of the Newtonian node precessions of L and \textcolor{black}{LR 2}  is far from vanishing, given the actual orbital configurations of both the satellites. Section\,\ref{sec3} is dedicated to the impact of the orbital injection errors on the uncancelled sum of the satellites' node shifts, while in  Section\,\ref{sec4} the consequences of the imperfect knowledge of the Earth's angular momentum on the proposed measurement are treated. A major drawback common to all the analyses performed so far by Ciufolini and coworkers is pointed out in Section\,\ref{sec5}. Section\,\ref{sec6} resumes the findings of the paper, and conclusions are offered.
\section{How accurate is the \textcolor{black}{cancellation} of the effect of the Earth's $J_2$ on the sum of the nodes?}\lb{sec2}
From \rfr{OLT} and \rfr{OJ2}, \textcolor{black}{one can derive} the following ratio of the sums of the classical to the relativistic node precessions of L and \textcolor{black}{LR 2}
\eqi
\mathcal{R}^{J_2}\lb{Erre}\doteq  \rp{\dot\Omega_{J_2}^\mathrm{L} + \dot\Omega_{J_2}^\mathrm{\textcolor{black}{LR2}}}{\dot\Omega_\mathrm{LT}^\mathrm{L} + \dot\Omega_\mathrm{LT}^\mathrm{\textcolor{black}{LR2}}}= -\rp{3\,c^2\,J_2\,R^2\,\sqrt{M}}{4\,\sqrt{G}\,S}\,\qua{
\rp{\rp{\cos I_\mathrm{L}}{a_\mathrm{L}^{7/2}\,\ton{1-e_\mathrm{L}^2}^2} + \rp{\cos I_\mathrm{\textcolor{black}{LR2}}}{a_\mathrm{\textcolor{black}{LR2}}^{7/2}\,\ton{1-e_\mathrm{\textcolor{black}{LR2}}^2}^2}}{\rp{1}{a_\mathrm{L}^3\,\ton{1 - e^2_\mathrm{L}}^{3/2}} + \rp{1}{a_\mathrm{\textcolor{black}{LR2}}^3\,\ton{1 - e^2_\mathrm{\textcolor{black}{LR2}}}^{3/2}}}
}.
\eqf

Its smallness provides a measure of how accurate a LT test can be by summing up the node rates of both the spacecraft; in principle, it should vanish giving an ideally clean measurement of the combined gravitomagnetic precessions.

The mean values of $a,\,e,\,I$ of L and \textcolor{black}{LR 2} over 127 days are \citep[Tab.\,1]{2023EPJC...83...87C}
\begin{align}
a_\mathrm{L} \lb{aL}& = 12\,270.020705\,\mathrm{km}, \\ \nonumber \\
a_\mathrm{\textcolor{black}{LR2}} \lb{aCL}& = 12\,266.1359395\,\mathrm{km}, \\ \nonumber \\
e_\mathrm{L} \lb{eL}& = 0.00403, \\ \nonumber \\
e_\mathrm{\textcolor{black}{LR2}} \lb{eCL}& = 0.00027, \\ \nonumber \\
I_\mathrm{L} \lb{IL}& = 109.8469\,\mathrm{deg}, \\ \nonumber \\
I_\mathrm{\textcolor{black}{LR2}} \lb{ICL}& = 70.1615\,\mathrm{deg}.
\end{align}

Concerning the removal of the competing effect of the Earth's $J_2$ from the sum of the nodes of L and \textcolor{black}{LR 2},
\citet[pp.\,4-5]{2023EPJC...83...87C} explicitly write:
\virg{The non-relativistic nodal drift of two satellites with supplementary
inclinations, and with the same semimajor axis,
is equal and opposite. So by properly adding the two secular
nodal shifts, it will be possible to eliminate the dominant classical
shift and to accurately measure the general relativistic
node shift. [\ldots] Hence by properly adding the
shift of the nodes of the two satellites we achieve a measurement
which is purely the frame dragging. [\ldots] For two satellites with supplementary inclinations,
the elimination of the bias due to the even zonals in the test of
frame-dragging is clearly shown by the well-known equation\footnote{It is \rfr{OJ2} of the present paper.}
for the secular nodal drift of a satellite due to the even zonals. [\ldots] the largest
nodal drift of an Earth satellite is by far due to the even zonal
harmonic of degree two, the Earth quadrupole moment. [\ldots] the supplementary inclination of configuration of LARES 2 and LAGEOS effectively allow
elimination of such a systematic bias.}

In fact, it seems not to be the case. Indeed, \rfr{Erre}, calculated with the values of \rfrs{aL}{ICL}, yields
\eqi
\left|\mathcal{R}^{J_2}\right| = 4\,918,
\eqf
that is, the sum of the nominal  node precessions due to the first even zonal harmonic $J_2$ of the geopotential is still almost $5\,000$ times larger than the sum of the theoretically predicted LT node precessions. It has a major impact on the evaluation of other systematic bias due to the errors in some of the parameters entering \rfr{Erre}, as it is demonstrated in the next Sections.
\section{The impact of the orbit injection errors}\lb{sec3}
A potentially major issue is represented by the impact of the orbit injection errors of \textcolor{black}{LR 2} (and of L as well) which would bias the combined LT rates through the Newtonian oblateness-driven precessions. According to \citet[Tab.\,3]{2023EPJC...83...87C}, it would amount to less than $0.1$ per cent of the combined LT effect.

Again, this does not appear to be the case.

The $J_2$-induced bias due to the orbit injection errors $\delta\mathcal{R}_\mathrm{inj}$ can be (optimistically) calculated as
\eqi
\delta\mathcal{R}_\mathrm{inj} = \sqrt{\sum_{\xi}\ton{\derp{\mathcal{R}^{J_2}}{\xi}\delta\xi}^2},\,
\xi = a_\mathrm{L},\,a_\mathrm{\textcolor{black}{LR2}},\,I_\mathrm{L},\,I_\mathrm{\textcolor{black}{LR2}},\,e_\mathrm{L},\,e_\mathrm{\textcolor{black}{LR2}},\lb{derre}
\eqf
where $\delta\xi$ represents some measure of the known error in the orbital element $\xi$; it is intended that \rfr{derre} must be calculated with the nominal value of $J_2$, not with its uncertainty $\delta J_2$.
From the number of significant digits quoted in \citet[Tab.\,1]{2023EPJC...83...87C} and reported in \rfrs{aL}{ICL}, it can be argued that the errors in the orbital elements of L and \textcolor{black}{LR 2} should be as follows:
\begin{align}
\delta a_\mathrm{L} \lb{daL}& \simeq 10^{-6}\,\mathrm{km} = 1\,\mathrm{mm}, \\ \nonumber \\
\delta a_\mathrm{\textcolor{black}{LR2}} \lb{daCL}& \simeq 10^{-7}\,\mathrm{km} = 0.1\,\mathrm{mm}, \\ \nonumber \\
\delta e_\mathrm{L} \lb{deL}& \simeq 10^{-5}, \\ \nonumber \\
\delta e_\mathrm{\textcolor{black}{LR2}} \lb{deCL}& \simeq 10^{-5}, \\ \nonumber \\
\delta I_\mathrm{L} \lb{dIL}& \simeq 10^{-4}\,\mathrm{deg} = 0.36\,\mathrm{arcsec}, \\ \nonumber \\
\delta I_\mathrm{\textcolor{black}{LR2}} \lb{dICL}& \simeq 10^{-4}\,\mathrm{deg} = 0.36\,\mathrm{arcsec}.
\end{align}

By calculating \rfr{derre} with \rfrs{deL}{deCL}, one obtains
\eqi
\delta\mathcal{R}^e_\mathrm{inj} \simeq 1.2,
\eqf
corresponding to a $\simeq 120$ per cent bias in the added LT rates; an improvement by 3 orders of magnitude in determining the satellites' eccentricities would be needed to reach the $0.1$ per cent level claimed by \citet{2023EPJC...83...87C}. An even worse case occurs for the inclinations. Indeed, \rfr{derre}, calculated with \rfrs{dIL}{dICL}, yields
\eqi
\delta\mathcal{R}^I_\mathrm{inj} \simeq 50.2,
\eqf
corresponding to a $\simeq 5\,000$ per cent error in the combined LT effect; the inclinations of the satellites' orbital planes should be known 4 orders of magnitude better than now, i.e. with an accuracy as good as
\eqi
\delta I\simeq 0.036\,\mathrm{milliarcseconds}\,(\mathrm{mas}) = 36\,\mathrm{microarcseconds}\,(\mu\mathrm{as}).
\eqf

In order to make a simple calculation about the possibility of actually improving \rfrs{deL}{dICL} by the required $3-4$ orders of magnitude over a reasonable time span $T$, it can be assumed that the uncertainty $\delta\kappa$ with which any of $\kappa = e_\mathrm{L},\,e_\mathrm{\textcolor{black}{LR2}},\,I_\mathrm{L},\,I_\mathrm{\textcolor{black}{LR2}}$ can be determined goes as $\simeq 1/\sqrt{N_T}$, where $N_T$ is the number of experimental determinations of $\kappa$ during the time span $T$.
Lets one assume to have calculated a value of $\kappa$ from the processed data at the end of every orbital arc, which is typically 7 days long for the satellites of the LAGEOS type. Thus, $N_\mathrm{yr}\simeq 52$ values of $\kappa$ would be obtained after 1 yr, corresponding to a reduction  of the error $\delta\kappa$ by a modest factor of $\sqrt{N_\mathrm{yr}}\simeq 7$. After 30 yr, $\delta\kappa$ would be reduced just by a factor of $\simeq \sqrt{30\times N_\mathrm{yr}} = 40$. It is clear that making the uncertainties of \rfrs{deL}{dICL} smaller by a factor of $1\,000-10\,000$ over any reasonable temporal interval is unrealistic.
\section{The impact of the uncertainty in the Earth's angular momentum}\lb{sec4}
Another major issue arising because of the too large value of \rfr{Erre} is the uncertainty $\delta S$ in the Earth's angular momentum  $S = \mathcal{I}\,\upomega$, where $\mathcal{I}$ is the moment of inertia (MoI) of our planet, and $\upomega$ is its angular speed.
Actually, to the knowledge of the present author, $\delta S$ is not explicitly reported anywhere in the literature; nonetheless, plausible guesses for it can be inferred from the significant digits with which some relevant related quantities are quoted.

As an example, Ren\,et\,al.\,\citep[p.\,2528]{2022ClDy...58.2525R} yield
\begin{align}
\mathcal{I} &\simeq 8.04\times 10^{37}\,\mathrm{kg\,m}^2,\\ \nonumber \\
\upomega &\simeq 7.29\times 10^{-5}\,\mathrm{s}^{-1},
\end{align}
implying an uncertainty in the MoI of the order of
\eqi
\delta\mathcal{I} \simeq 10^{35}\,\mathrm{kg\,m}^2.
\eqf

Thus, it can be inferred
\begin{align}
S &\simeq 5.86\times 10^{33}\,\mathrm{kg\,m}^2\,\mathrm{s}^{-1}, \\ \nonumber \\
\delta S &\simeq 7.3\times 10^{30}\,\mathrm{kg\,m}^2\,\mathrm{s}^{-1},
\end{align}
corresponding to a relative uncertainty
\eqi
\rp{\delta S}{S} \simeq 0.0012. \lb{dSS1}
\eqf

Another way to attack the problem consists in looking at the Earth's angular momentum per unit mass $J$ \citep[p.\,156]{iers10}
\eqi
J \simeq 9.8\times 10^8\,\mathrm{m}^2\,\mathrm{s}^{-1}.
\eqf
Thus, it can be guessed
\eqi
\delta J\simeq 10^7\,\mathrm{m}^2\,\mathrm{s}^{-1},
\eqf
yielding
\eqi
\rp{\delta J}{J} = \rp{\delta S}{S}\simeq 0.010.\lb{dSS2}
\eqf

Since $\mathcal{I} = \alpha\,M\,R^2$, where $\alpha$ is the normalized moment of inertia (nMoI), or moment of inertia factor, the uncertainty in $S$ can be inferred also from that of $\alpha$.
According to the NASA Space Science Data Coordinated Archive (NSSDC),
 it is\footnote{See \url{https://nssdc.gsfc.nasa.gov/planetary/factsheet/earthfact.html} on the internet.}
 \eqi
 \alpha = 0.3308.
 \eqf
 
Thus, by assuming
\eqi
\delta\alpha\simeq 0.0001,
\eqf
one gets
\eqi
\rp{\delta\alpha}{\alpha} = \rp{\delta S}{S}\simeq 0.0003.\lb{dSS3}
\eqf

According to \rfr{Erre}, the systematic bias of $J_2$ in the combined LT node precessions due to the uncertainty in $S$ is given by
\eqi
\delta\mathcal{R}_S = -\mathcal{R}^{J_2}\,\rp{\delta S}{S};\lb{dRS}
\eqf
if the Newtonian oblateness-driven precessions canceled out to a sufficiently accurate level, $\delta\mathcal{R}_S$ would be negligible.
Instead, the previously inferred guesses for the relative uncertainty in $S$, applied to \rfr{dRS}, yield quite large figures. Indeed, \rfr{dSS1} returns
\eqi
\delta\mathcal{R}_S \simeq 5.9,
\eqf
corresponding to a systematic error in the LT combined signal as large as 590 per cent.
The bias corresponding to \rfr{dSS2} is even worse, amounting to
\eqi
\delta\mathcal{R}_S \simeq 49.2,
\eqf
implying a staggering $4\,918$ per cent error in the added LT precessions.
Instead, \rfr{dSS3} yields
\eqi
\delta\mathcal{R}_S \simeq 1.5,
\eqf
which corresponds to a systematic uncertainty in the LT signature of about 150 per cent.
\section{Why \textcolor{black}{was} the gravitomagnetic field of the Earth neither modeled nor solved for?}\lb{sec5}
From a practical point of view, the LT effect would be nothing more than one of the many other dynamic features of various origins entering the equations of motion of the satellites and whose characteristic parameter(s) are to be estimated in the data reductions along with those describing the behaviour of measuring devices, the propagation of electromagnetic waves, the spacecraft's state vector, etc. Indeed, the common practice in satellite geodesy, astrodynamics and astronomy is that, if one wants to put to the test a certain dynamical feature X she/he is interested in, she/he must nothing else to do than explicitly modeling it along with the rest of known dynamics and other pieces of the measurement process, and simultaneously estimating one or more parameters characterizing it along with many other ones taking into account for other accelerations, etc., and inspecting the resulting covariance matrix to look at their mutual correlations. Instead, inspecting some sort of \virg{spurious} residuals constructed without including X in the models fit to a given set  of observations  is not a correct procedure since a possible signature with almost the same features of the expected one may be due just to some fortunate partial mutual \textcolor{black}{cancellation} of other effects having nothing to do with X itself. Furthermore, X may partly or totally be absorbed in the estimated values of other parameters solved for in the data reduction. In other words, the gravitomagnetic field of the Earth should be simultaneously estimated along with all the other coefficients characterizing the geopotential by using the same data sets which may be varied from time to time by their extension, starting date, and type of observations.

On the contrary, for reasons unknown to the present author, nothing of this has ever been done:
the Earth's gravitomagnetic field has never been modeled, and no dedicated parameter(s) have ever been estimated, producing just time series of post-fit residuals of the satellites' nodes\footnote{They are not even directly measurable quantities.}, allegedly accounting in full for the  unmodeled  dynamics which includes the LT acceleratio\textcolor{black}{n} as well. Another puzzling issue is that there are several SLR stations scattered around the globe \citep{2002AdSpR..30..135P} where skilled teams of space geodesists routinely process laser ranging data from so many geodetic satellites with several dedicated softwares \citep{2017G&G.....8..213E}; yet, despite this, no one has ever tried to (correctly) perform LT tests independently of Ciufolini, or, if anyone has done so, she/he has not made her/his results public in the peer-reviewed literature. There are just some conference proceedings \citep{Riesetal03a,Riesetal03b,2008lara.workE..19R,2009IAU...261.1402R} in which
the author did not model and estimate the LT acceleration either. The same holds also for a few independent studies recently published in peer-reviewed journals by former coworkers of Ciufolini \citep{2019Univ....5..141L,2020Univ....6..139L}. In principle, there should be nothing easier for so many competent and expert people worldwide than adding one more acceleration in the data reduction softwares and estimating  one more parameter.
\section{Summary and conclusions}\lb{sec6}
The orbital parameters of the newly launched laser-ranged geodetic satellite \textcolor{black}{LARES 2} do not allow to cancel out the sum of \textcolor{black}{its} classical oblateness-driven node precessions  and \textcolor{black}{those} of its cousin LAGEOS; indeed, it still amounts to about $5\,000$ times the sum of the LT node rates.

This fact has relevant consequences on the overall systematic bias induced by the Earth's quadrupole moment due to the errors in other parameters entering the ratio of the sum of the precessions due to it to the sum of the LT ones.

The orbital injection errors in the eccentricities and the inclinations of the two satellites yields a $\simeq 5\,000$ per cent systematic error in the combined LT signature. Improvements by $3-4$ orders of magnitude in the determination of such orbital elements, unlikely to be attainable over any realistic time span, would be required.

Also the imperfect knowledge in the Earth's angular momentum has a relevant impact on the ratio of the classical to relativistic summed precessions; depending on how its uncertainty is assessed, the systematic bias induced on the LT signal  ranges from $150$ to $5\,000$ per cent.

Recurring to the formalism of the Keplerian orbital elements may be useful to do preliminary sensitivity analyses and to suitably design experiments. Actual tests must, instead, be performed by modeling the dynamical feature one is interested in, and by estimating its characteristic parameter(s) along with the other ones taking into account the rest of dynamics in fitting dynamical and measurement models to the observations.
\textcolor{black}{As of now, this standard procedure has not yet been implemented, though it is} common to all data reductions in satellite geodesy, geodynamics, astrodynamics and astronomy.
\section*{Acknowledgements}
 I gratefully thank two referees for their constructive remarks.
\bibliography{ebbastabib}{}
\end{document}